\documentclass[footinbib,twocolumn,english,aps,prb,showpacs,superscriptaddress,floats,amsmath,amssymb,floatfix]{revtex4-1}

\usepackage[latin9]{inputenc}
\usepackage{color}
\usepackage{amsmath}
\usepackage{graphicx}
\usepackage{graphics}
\usepackage{gensymb}
\usepackage[caption=false]{subfig}
\usepackage{pdftexcmds}
\usepackage{ifpdf}
\usepackage{amssymb}
\usepackage{subfig}
\usepackage{dsfont}

\makeatletter
\@ifundefined{textcolor}{}
{%
 \definecolor{BLACK}{gray}{0}
 \definecolor{WHITE}{gray}{1}
 \definecolor{RED}{rgb}{1,0,0}
 \definecolor{GREEN}{rgb}{0,1,0}
 \definecolor{GREEN2}{rgb}{0,0.4,0}
 \definecolor{BLUE}{rgb}{0,0,1}
 \definecolor{CYAN}{cmyk}{1,0,0,0}
 \definecolor{MAGENTA}{cmyk}{0,1,0,0}
 \definecolor{YELLOW}{cmyk}{0,0,1,0}
 \definecolor{YELLOW2}{cmyk}{0,0,1,0.6}
 \definecolor{ORANGE}{rgb}{1,0.22,0}

 }

\def \nn {\nonumber}
\def \nn{\nonumber}
\def \be {\begin{equation}}
\def \ee {\end{equation}}
\def \bea {\begin{eqnarray}}
\def \eea {\end{eqnarray}}

\allowdisplaybreaks
\makeatother

\begin{document}

\title{Vortex lattice in two-dimensional chiral XY ferromagnets and inverse
    Berezinskii-Kosterlitz-Thouless transition}

 \author{Alejo Costa Duran}
 \affiliation{Facultad de Ciencias Exactas, Universidad Nacional de La Plata, C.C. 67, 1900 La Plata, Argentina}
 \affiliation{Instituto de F\'isica de L\'iquidos y Sistemas Biol\'ogicos, CCT La Plata, CONICET}
 \author{ and Mauricio Sturla}
\affiliation{Facultad de Ciencias Exactas, Universidad Nacional de La Plata, C.C. 67, 1900 La Plata, Argentina}
\affiliation{Instituto de F\'isica de L\'iquidos y Sistemas Biol\'ogicos, CCT La Plata, CONICET}

\begin{abstract}
 In this Letter we will show that, in the presence of a properly modulated Dzyaloshinskii-Moriya (DM) interaction, a $U(1)$ vortex-antivortex lattice appears at low temperatures for a wide range of the DM interaction. Even more, in the region dominated by the exchange interaction, a standard BKT transition occurs. In the opposite regime, the one dominated by the DM interaction, a kind of inverse BKT transition (iBKT) takes place. As temperature rises, the vortex-antivortex lattice starts melting by annihilation of pairs of vortex-antivortex, in a sort of ``inverse'' BKT transition.
\end{abstract}
\maketitle

{\it Motivation.--}

Since the seminal works from Berezinskii, Kosterlitz and Tholues \cite{berezinsky1970,berezinsky1972,kosterlitz1973}, 
the BKT-transition (a topological defect mediated phase transition) and the existence of the $U(1)$-vortices in the disordered phase of the two dimensional Heisenberg XY-ferromagnets, have disruptively influenced the physics of condensed matter, giving topology a central role in the physics beyond the Landau paradigm.
The existence of periodic arrangements of these topologically singular  excitations ($U(1)$-vortices), on the other hand, is also very well established both from theoretical and experimental points of view, and have played a central role in condensed matter since they were postulated by Abrikosov 1957 \cite{abrikosov_57}, and observed in type II superconductors,  ten years later by Essmann and Tr\"auble \cite{essmann1967}. These $U(1)$-vortices lattices appear in many different materials, form HTC superconductors to $^4$He superfluids, and BEC's, or in the so called fully frustrated XY-model (FFXY) used to model periodic arrays of Josephson junctions \cite{yarmchuk1979,madison2000,abo2001,teitel_83a, teitel_83b, teitel2013}. Nevertheless, the existence of such lattices as stable states of 2-dimensional pure  magnetic materials, is not so well known.

In the past years, a new kind of magnetic materials, known as chiral magnets, has called the attention of the community of condensed matter, due to their capability for supporting periodic arrangements of another type of topologically non-trivial magnetic excitation. In this case, a smooth kind of topological excitation named skyrmions, relevant for memory devices and quantum computing technology. The key ingredient to stabilize these lattices seems to be chirality. In these magnets, it is widely assumed that this chirality is a consequence of an antisymmetric exchange interaction, the Dzyaloshinskii-Moriya (DM) interaction, originated in the spin-orbit coupling of non-centrosymmetric magnetic materials. The technological implication of these chiral magnets, and in particular of the skyrmion crystal phases they support, has motivated a race for the enhancement and modulation of the DM interaction by different methods, leading to the emergence of a new research field named ``spin-orbitronics''\cite{yang18}. Recent studies show that in carefully designed heterostructures of chiral magnets, and by proper application of electric fields, among other techniques, it is possible to achieve DM interactions of the same order of magnitude that the exchange one\cite{yang18,nembach15,luo19}. Also, it has been shown that the DM magnitude could grow linearly with the applied electric fields and can also be modulated\cite{yang18}, opening new technological possibilities.

In this context, we will revisit the XY models for ferromagnets, now in the presence of strong DM interactions. We will show in this Letter that for values of the DM interaction slightly stronger than the exchange interaction, a vortex-antivortex lattice can be stabilised at low temperatures. Even more surprisingly, in the region dominated by the DM interaction, the system undergoes a finite temperature phase transition in the same universality class than the BKT transition. By mapping the system to a 2D-Coulomb gas, we interpret this transition as a sort of inverse BKT transition (iBKT), in which the vortex lattice starts melting, as temperature rises, by annihilation of vortex-antivortex pairs. In what follows, we derive the results leading to this conclusion.\\

{\it Model.--} We will start by considering a ferromagnetic XY-Hamiltonian in the presence of the antisymmetric DM interaction on a square two-dimensional lattice:
\be
\mathcal{H}=-\sum_{\mathbf{r}_i,\hat{\mu}}J_{i\hat{\mu}}\mathbf{S_{i}}\cdot \mathbf{S_{i+\hat{\mu}}}+\mathbf{D_{i\hat{\mu}}}\cdot\left( \mathbf{S_{i}}\times \mathbf{S_{i+\hat{\mu}}} \right).\label{eq:XYDMNotacionVecinos}
\ee
Where $\hat{\mu}$ represents the unit vectors along positive axis directions, the spin $\mathbf{S}$ is a two component unimodular vector, $J>0$ is the ferromagnetic exchange coupling and the $\mathbf{D}_{i\hat{\mu}}$ vectors pointing outside the plane of the lattice (let say the XY-plane) represent the DM interaction. 

\begin{figure}[htb]
\includegraphics[width=8.7cm]{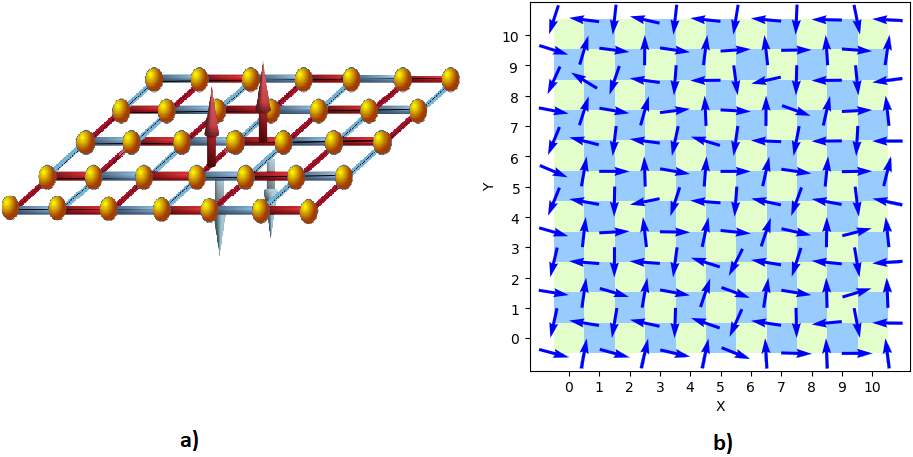}
\caption{Figure a) shows the direction of the $D$ vectors, represented by red and light blue out of plane arrows of a given plaquette, when circulating the lattice in the positive direction of the axes. The colours of the bonds indicate the corresponding directions for the remaining $D$-vectors. Figure b) shows a vortex lattice of $10 \times 10$ spins obtained through a standard Metropolis Monte Carlo method, the $X$ and $Y$ axes represent the direction on the lattice in units of the lattice spacing. Green plaquettes hold a counter-clockwise vortex while blue plaquettes hold a clockwise vortex and spins in each sites are represented with blue arrows. }  
\label{fig:Ds}
\end{figure}

\begin{figure}[!]
\centering
\parbox{8cm}{
  \includegraphics[width=8cm]{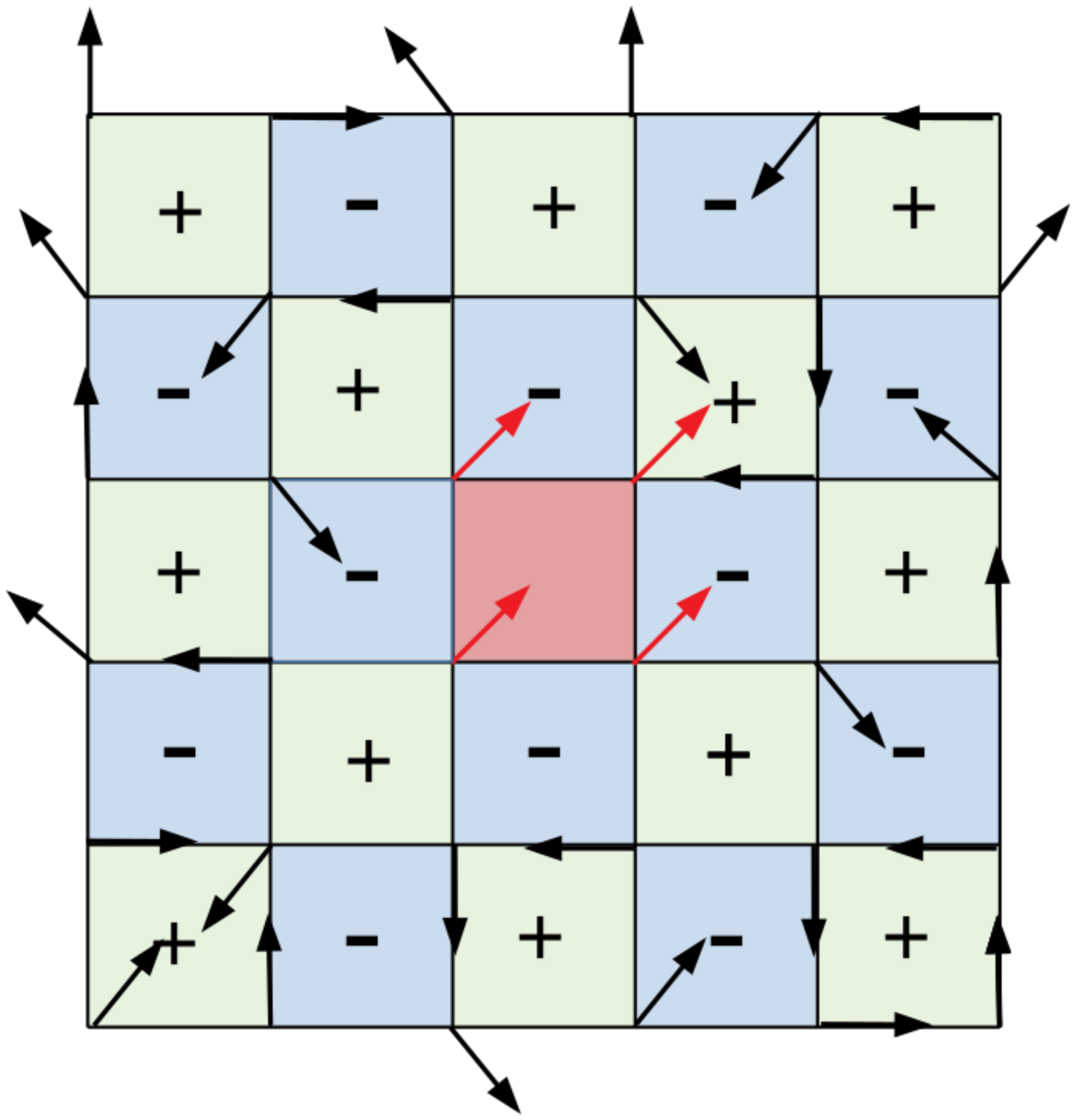}
\caption{Illustrative: The figure shows a configuration in which only one vortex, a positive one, has been eliminated from the lattice, and the remaining vortex-antivortex lattice has not been affected; $D$ dominated regime.}
\label{fig:11}}
\end{figure}

We define  new variables $\varphi_{i, \hat{\mu}}$ and $\mathcal{J}_{i, \hat{\mu}}$, in terms of which the original variables read $J_{i, \hat{\mu}}=\cos(\varphi_{i, \hat{\mu}})\mathcal{J}_{i, \hat{\mu}}$
and $D_{i, \hat{\mu}}=\sin(\varphi_{i, \hat{\mu}})\mathcal{J}_{i,\hat{\mu}}$, and the Hamiltonian can be recast in the following way:  

\be
\mathcal{H}=-\sum_{\mathbf{r_i},\hat{\mu}} \mathcal{J}_{i,\hat{\mu}} \cos(\theta_i-\theta_{i+\hat{\mu}}-\varphi_{i,\hat{\mu}}),
\label{eq:XYDMMapeoFFXY}
\ee
where $\theta_i$ represents the angle with respect to a given fixed direction of the $\mathbf{S}_i$ vector. This Hamiltonian has been previously studied by Teitel and collaborators in the context of Josephson junction arrays (see for example Ref. \onlinecite{teitel2013} and references therein), showing for the first time that it supports a vortex lattice at low temperatures. The $\varphi$ configuration that will be studied here explicitly breaks the $\mathbb{Z}_2$-symmetry present in the models studied by Teitel, and the phenomenology derived from it, as far as we know, has not been previously reported. \footnote{The FFXY models studied in these works, have an uniform flux on each plaquette with $f_c=\frac{\phi_c}{\phi_0}$, where $\phi_0$ is the flux quanta and $\phi_c$ is the flux on each plaquette, constant all along the sample. In the case studied here, there is no such flux, of course, but the analogy is immediate, and it would corresponds to an alternating flux through neighbouring plaquettes. So that, the $\mathbb{Z}_2$ symmetry that in the FFXY model is spostaneously broken, is explicitly broken here} The stable configurations, of course,  will depend on the particular field configuration $\varphi_{i,\hat{\mu}}$ chosen. A simple non-trivial choice for $\varphi_{i,\hat{\mu}}$ corresponds to a constant value, $|\varphi_{i,\hat{\mu}}|=\varphi$, with alternating signs along the bonds, as depicted in Fig. \ref{fig:Ds}a). 
This $\varphi$ configuration, the only one that we consider here\footnote{By proper modulation of DM interaction, lattices of vortices of arbitrary sizes can be built with the same tessellation technique}, leads to a lattice of {\it minimal vortices} \footnote{In this discrete context we will call a vortex, centered in a given plaquette, to every configuration of the order parameter $\theta(\mathbf{r_i})$ that, when circulating around the center of such plaquette, accumulates an integer times of $2\pi$'s, in steps smaller than $\pi$. A minimal vortex, then, is a vortex of the minimal size, i.e. a vortex of the size of the plaquette. This kind of vortices has been considered widely in the literature (see, for example, Ref.~\onlinecite{teitel2013} and references therein).}.
The energy condition imposed by Hamiltonian \eqref{eq:XYDMMapeoFFXY} for a given plaquette with $D$-vectors pointing down along the lower and right bonds, and pointing up in the other two bonds reads:
\bea
0&=& \sin(\frac{2\theta_1-\theta_2}{2})\cos(\frac{\theta_2-2\varphi}{2})\\ \nn
0&=& 2\sin(\frac{2\theta_2-\theta_1-\theta_3}{2})\cos(\frac{\theta_3-\theta_1-2\varphi}{2})\\ \nn
0&=& 2\sin(\frac{2\theta_3-\theta_2}{2})\cos(\frac{\theta_2+2\varphi}{2}),
\label{minimal_energy}
\eea
where a possible global phase has been set to zero, because of $U(1)$ global invariance. 
The angles $\theta_1,..., \theta_4$ are numerated counterclockwise starting at the lower left corner of the plaquette. Ferromagnetic and counterclockwise vortex configurations with $\Delta \theta_{i+1,i}=\pi/2$ satisfy the condition (\ref{minimal_energy}) for any value of $\varphi$, with corresponding energies:
\begin{subequations}
	\begin{align}
		E_f &= -4\cos(\varphi), \\
		E_v &= -4\cos(\pi/2-\varphi).
	\end{align}
        \label{energies}
\end{subequations}
It is straightforward to check that the four adjacent plaquettes (two corners sharing) to the one considered, have the same both trivial and non-trivial solutions with the same energies, but with a clockwise vortex instead of anticlockwise. That means that, for $\varphi \in [0,\frac \pi 4)$ the ground state becomes ferromagnetic, while for $\varphi \in (\frac \pi 4, \pi/2]$ a vortex-antivortex lattice is stabilised, as in Fig \ref{fig:Ds}b). The high symmetry in Fig \ref{fig:Ds}b), can lead us to mistakenly conclude that a plaquette surrounded by four plaquettes with vortices of one type, only admits a vortex of the opposite type. The illustrative Fig \ref{fig:11} can clarify this aspect. Finally, for $\varphi=\frac\pi 4$ the possibility of a coexistence of both phases can not be discarded \footnote{Both configurations have the same $U(1)$ manifold degeneracy}.\\

{\it Low temperature effective theory.--}
We will start the analysis of the system by performing a low temperature expansion following Savit \cite{savit78}. We notice that  $\mathcal{J}_{i,\hat{\mu}}$ is independent of bond and lattice site, 
and representing $\varphi_{i,\hat{\mu}}$ as a vector $\boldsymbol{\varphi}_i$ with components $\varphi_{\hat{x},i} = (-1)^{x_i+y_i}\varphi$ and $\varphi_{\hat{y},i} = (-1)^{x_i+y_i+1}\varphi$ on each site $i$, the partition function  associated with the Hamiltonian (\ref{eq:XYDMMapeoFFXY}) can be written as: 
\begin{equation}
		Z = \int_{-\pi}^{\pi} \prod_{j} \frac{d\theta_j}{2\pi} \exp \left[ \beta \mathcal{J} \sum_{i,\mu} \cos(\theta_i-\theta_{i+\mu}-\boldsymbol{\varphi}_i\cdot\hat{\mu}) \right]
\end{equation}

Expanding each exponential in series of Bessel functions \cite{abramowitz48}, this partition function can be recast as:
\begin{equation}\label{ParticionPreIntegracion}
	    \begin{split}
		    Z = &\sum_{\left\{ n \right\}} \left( \prod_{i,\mu} I_{n_{i,\mu}}(\beta \mathcal{J}) \exp \left[-in_{i,\mu}\boldsymbol{\varphi}_i\cdot\hat{\mu} \right]  \right)\\
		    &\int_{-\pi}^{\pi} \left( \prod_{j} \frac{d\theta_j}{2\pi} \right) \exp \left[ \sum_{i,\mu} in_{i,\mu}(\theta_i-\theta_{i+\mu}) \right],
		\end{split}
\end{equation}
where $\{n\}$ represents a sum over all possible integers configurations, one ${n_{i,\mu}}$ per bond, and  $I_n(\beta \mathcal{J})$ are the modified Bessel functions of first kind of order $n$. In this factorised way, integration over each angular variable can be done, and a theory on the discrete variable $n$, with the condition:
\begin{equation}
		\boldsymbol{\Delta} \cdot \mathbf{n}_i = n_{i,x}-n_{i-\hat{x},x}+n_{i,y}-n_{i-\hat{y},y}=0, 
\end{equation}
is obtained. Of course, this null discrete divergence condition can be immediately fulfilled by a discrete rotor $n_{j,\mu} = \epsilon_{\mu \nu} \Delta_\nu \phi_j$, where $\left\{ \phi \right\}$ is a set of integers defined on the dual lattice, that it is the square lattice formed by the centre of the original plaquettes.
Introducing Dirac's deltas, $\sum_{k=-\infty}^{\infty} \delta(\phi -k) = \sum_{m=-\infty}^{\infty} e^{i2\pi m_j \phi_j}$, the sum over discrete variables can be turned into integrals of now continuous $\phi_j$ and, at sufficient low temperature, the low energy partition function can be written as:
\begin{equation}\label{ParticionTBaja}
		Z = \int \mathcal{D}\phi \sum_{ \left\{ m \right\}} \exp \left[ \sum_{\mu,j} -\frac{1}{2\beta \mathcal{J}} (\Delta_\mu \phi_j)^2 + i2\pi M_j \phi_j \right],
	\end{equation}
where $M_j = m_j - (-1)^{x_j+y_j}\frac{2\varphi}{\pi}$ has been introduced.


Performing the Gaussian integrals by Fourier transforming the fields, the partition function reads:	    
\begin{equation}\label{PartitionFunctionVortex}
		Z = Z^{0} \sum_{\left\{ m \right\}} \exp \left[ -\frac{\beta\mathcal{J}}{8}\sum_{i,j}M_iV_{ij}M_j \right].
\end{equation}

where:
\bea
Z^{0}& =& \exp \left[ -\int_{-\pi}^{\pi}d^2q \frac{1}{2} \ln \left( \frac{2K(q)}{\pi} \right) \right]\\ \nn
V_{ij} &= &\int_{-\pi}^{\pi} d^2q \frac{e^{i\vec{q}(\vec{i}-\vec{j})}}{\left( 1-\frac{1}{2}\sum_{\mu}\cos(\vec{q}\cdot\hat{\mu})\right)},\nn
\eea
with $K(q)$ approximated by
\be
K(q)=\frac{1}{2\beta \mathcal{J}\pi^2} ( 1-\frac{1}{2}\sum_{\mu}\cos(\vec{q}\cdot\hat{\mu})).\nn
\ee

For small $|q|$ the potential reduces to the Coulomb gas potential, $V_{ij} \simeq \int_{-\pi}^{\pi} 2\frac{e^{iq(i-j)}}{|q|^2}d^2q$, that after proper regularisation by imposing charge neutrality $\sum_{i}M_i=0\;$ \cite{schakel}, leads to the low temperature partition function:
\bea
Z &=& Z^{0} \sum_{\left\{ m \right\}} \exp \left[\beta\mathcal{J}\pi \sum_{i,j} M_i \ln(\left|\mathbf{R}_i-\mathbf{R}_j\right|)M_j\right. \nn \\ 
  &-&\left. \beta\mathcal{J}\pi \sum_{l}\left(\frac{1}{2} \ln(8)+\gamma \right) M_l^2 \vphantom{-\frac{\pi}{2T_1(\beta \mathcal{J})}}\right], 
\label{PartitionExplicita}
\eea
where $\gamma$ is the Euler-Mascheroni constant.

This low temperature effective theory is, in fact, an extension of the well known description of the XY-model, ($\varphi=0$), as a two-dimensional Coulomb gas \cite{savit78,jose1977renormalization}.

It is clear now that the system at low temperatures can be understood as a neutral Coulomb gas of excitations of charges $M$, whatever the value of $\varphi$ is. More precisely, at each value of $\varphi$ the ground state of the system corresponds to a configuration in which the charges $M$ take their minimal possible absolute value. It is interesting to note that, both in pure exchange regime and in pure DM regime, the charges become integer and the minimal possible values for $M$ correspond to $M_i=0, \forall i$. In the pure exchange regime, $\varphi \rightarrow 0$, the charge $M_i=m_i$, where $m_i$ represents the $i$-th topological charge in the low temperature theory \cite{savit78}. The condition $M_i=0, \forall i$, implies that no topological excitation is present at sufficient low temperature, as expected for a ferromagnetic ground state, Eq. \eqref{minimal_energy}.
On the other hand, in the pure DM regime, $\varphi \rightarrow \frac \pi 2$, the charge becomes $M_i=m_i-(-1)^{x_i+y_i}$. The condition $M_i=0, \forall i$, implies that $m_i=(-1)^{x_i+y_i}$ at each site, and a fully populated vortex-antivortex lattice emerge at sufficient low temperature, again as expected from the microscopical theory, Eq. \eqref{minimal_energy}. At any intermediate value of $\varphi$, the condition that $|M|$ must be the minimal possible shows that the ferromagnetic background  extends to all the region dominated by the exchange interaction, and the vortex-antivortex lattice background extends to all the region dominated by the DM interaction, also as predicted by the microscopic theory. The relevance of the effective theory relies on the interpretation of the excitations at low temperature. The study of the excitations in the microscopical theory could be very cumbersome, and the effective theory can shed some light on this matter. Excitations correspond to values of the charges different from their minimal values, and behave as a neutral Coulomb gas. On the exchange dominated regime, the minimal energy excitation corresponds to one pair of non minimal charges $M=\pm \mu$, which implies that a pair of one vortex and one antivortex has been created, i.e. $m=\pm 1$, and the well known phenomenology of the XY model follows. Very interesting features not present in the standard XY model appear for $\varphi < \pi/4$, but they will be discussed in a forthcoming paper. In this Letter we will discuss the phenomenology in the DM dominated regime.\\  

{\it DM dominated regime and Helicity modulus.--} 
As we have already mentioned, in the DM dominated regime the ground state corresponds to a regular arrangement of  vortices and antivortices, as depicted in Fig. \ref{fig:Ds}b). This can be seen from the minimal charge condition for $M_j = m_j - (-1)^{x_j+y_j}\frac{2\varphi}{\pi}$,  for $\frac{2\varphi}{\pi} > \frac 1 2$. No matter the value of $\varphi > \pi/4 $, the minimal condition is achieved by $m_j = (-1)^{x_j+y_j}$. Again, the minimal excitation is given by one pair of non-minimal opposite charges $M = \pm \mu$, but in this case it corresponds to $m=0$ in both charges \footnote{Since the system only presents minimal vortices, $|n|>1$ can not be distinguished from $|n|=1$ and excitations with $|n|>1$ have no physical meaning and are not considered. Nevertheless, numerically we observe that as temperature rises, vortex density only decrease}. That is to say, the minimal excitation corresponds to a pair of opposite charges that now have trivial winding number, one where before there was a vortex, and another one where before there was an anti-vortex. Or rephrasing, the excitation corresponds to the annihilation of a pair of a vortex and an antivortex! This conclusion is not easy to reach from the microscopical theory since, although difficult, configurations with only one vortex annihilated can be constructed (see illustrative Fig. \ref{fig:11}). On the effective theory, on the other hand, it is an immediate conclusion from the neutrality charge. We notice that, the ``effective charge'' oscillates with the position, in such a way that it is not possible to move only one charge without violating charge neutrality. 

As temperature starts to rise, more pairs of opposite charges are created (more pairs of vortex antivortex are annihilated) and eventually, at some temperature, they could decouple and decorrelate the systems much in the same way as the vortex do in the XY-model. We remark here that, although the transition shares many aspects with the BKT transition, the melting of the lattice goes in a direction inverse to the one in the BKT transition. In the present case, the system goes from topologically non-trivial entities in the stable state, to a decorrelated state dominated by topologically trivial excitations, so that we call this transition inverse BKT transition (iBKT). In order to support this picture, we compute the helicity modulus, as it is standard in the BKT transition, and show that the iBKT transition has the same universal jump that the standard BKT transition, and numerically show that vortex are annihilated by pairs. Introduction of a $\lambda_0$ long-wavelength ``twist'' on the local order parameter, with $k_0=2\pi/\lambda_0$, should rise the free energy by $O(k_0^2)$ over their ground state value, if the system is correlated, and should have no appreciable effect if the system is not\cite{ohta1979xy}. That is to say that the helicity modulus $\Upsilon$ \cite{fisher1973helicity}

\begin{equation}
    \Upsilon \equiv \left. \frac{\partial^2 F(T,k_0)}{\partial k_0^2}\right|_{k_0=0},
    \label{helicity}
\end{equation}
where $F(T,k_0)=-T\ln(Z(\Phi,T))/N$ is the free energy per unit volume, must be finite if the system is in the correlated phase and zero if it is not. In fact, BKT transition is characterised by a finite jump in $\Upsilon/T_c$ of $2/\pi$. In what follows we will show the numerical results for the helicity \eqref{helicity} for all values of the couplings and theoretical computations of the helicity modulus, following Ohta and Jasnow \cite{ohta1979xy} (see supplementary information), for extreme values of the parameters.\\   

\begin{figure}[!]
\centering
\parbox{8.6cm}{
  \includegraphics[width=7cm]{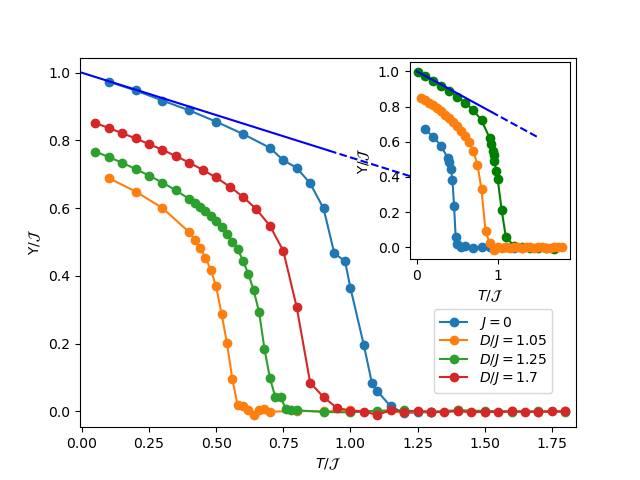}
\caption{The BKT behavior of $\Upsilon(T)$ for chiral XY model in the $D$ dominated regime is shown. A jump in $\Upsilon$ is appreciated at each value of $\mathcal{J}$, the theoretical prediction for $J\rightarrow 0$ is shown in purple. In the inset, the characteristic XY $\Upsilon(T)$ behavior for $J$ dominated regime is also shown, the curves with lower intercept correspond to lower values of $D/J$.}
\label{fig:2}}
\qquad
\begin{minipage}{8,6cm}
  \includegraphics[width=7cm]{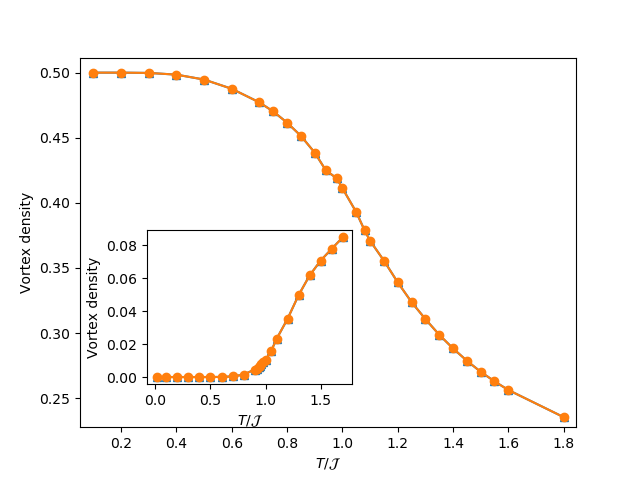}
\caption{The picture shows vortices and antivortices densities as temperature rises, for $D\gg J$. In the inset, the same densities show that standard phenomenology for $ D=0$}
 \label{fig:3}
\end{minipage}
\end{figure}

{\it Helicity and phase transition.--}
The Kosterlitz-Thouless renormalization group equations show that the helicity modulus $\Upsilon$ of a system of infinite size has a universal jump from the value $(2/\pi)T_c$ to zero at the critical temperature $T_c$. In Fig. \ref{fig:2}, the behavior of $\Upsilon$ as function of the temperature is shown for different values of $D/J$, where an abrupt jump in the helicity modulus at sufficient high temperature is observed. These results were obtained by a standard Metropolis Monte Carlo method with periodic boundary conditions on a square lattice of $32\times32$ sites. For extreme values, the theoretical prediction of the helicity in the correlated phase is also shown. We also compute numerically, positive and negative vortex density for different values of $D/J$ and we observe that, as temperature rises, both densities decrease at the same time, which implies that the vortex are annhilated by pairs. In Fig. \ref{fig:3}, both densities as a function of the temperature are shown for $J=0$. At each calculated temperature the densities have the same value. This is a non-trivial numerical result, that coincides with the neutrality charge condition of the effective theory and rules out the possibility depicted in Fig. \ref{fig:11}. The behavior of $\Upsilon$ is the one qualitatively expected for a BKT transition. The softening observed in the figure is due to finite size effect of the sample. Using the solution to the Kosterlitz-Thouless renormalization group equations and the 2D-Coulomb gas duality, it has been shown that BKT transitions obey a particular scaling law with the sample size, that allows us to determine the transition temperature $T_c$ \cite{weber1988monte}:
\bea \label{HelicityScaling}
\frac{\Upsilon(N,T)}{T\mathcal{J}}=\frac{\Upsilon^{\infty}(T)}{T\mathcal{J}}\left(1+\frac{1}{2} \frac{1}{\ln(N)+C}\right),
\eea
 where $\Upsilon(N,T)$ is the helicity modulus of the square lattice of $N$ sites, $C$ is an undetermined constant, and $\Upsilon^{\infty}(T)$ is the helicity modulus in the limit of $N\rightarrow \infty$. If the system undergoes a Kosterlitz-Thouless transition at a temperature $T_c$, we should obtain $\Upsilon^{\infty}(T_c)/(\mathcal{J}T_c)=2/\pi$. 
 
 For the determination of $T_c$ in the $D\gg J$ case, we follow the strategy developed by Weber and Minnhagen \cite{weber1988monte}.  We calculate $\Upsilon(N,T)$ for lattice sizes ranging from $32\times32$ to $128\times128$ and temperatures ranging from $T=0.885D$ to $T=0.91D$. For a given $T$, we make a least squares fit of $\Upsilon(N,T)/T$ to \eqref{HelicityScaling}. We find that the quantity $\Upsilon^{\infty}/T$ lies in the interval $0.61<\Upsilon^{\infty}/T<0.65$ for $0.885<T/D<0.9$. By this method $\Upsilon^{\infty}/T_c$ is determined to be $\Upsilon^{\infty}/T_c = 2/\pi \pm 0.03$ and we can estimate $T_c$ to be $T_c=0.892(8)D$.


\begin{figure}[!]
\includegraphics[width=.5\textwidth]{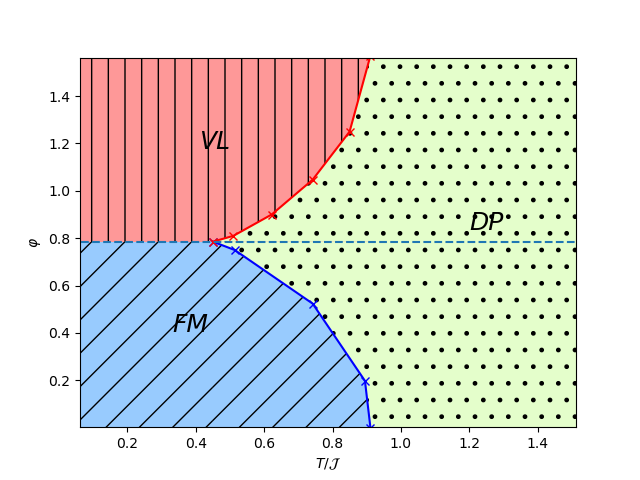}
\caption{The figure shows a qualitative phase diagram of the model studied. The vertical axis represents the variable $\varphi$ that goes from $\varphi=0$ (J dominated regime) up to $\varphi=\pi/4$ (D dominated regime). The value $\varphi=\pi/4$ (where $D=J$) is depicted with a dashed horizontal line. We identify three regions: the region VL where the system stays at a vortex lattice configuration; the FM where the system stays at a ferromagnetic configuration with quasi-long range order as in the standard XY model; and the decorrelated high temperature phase. As mentioned in previous sections, the possibility of a coexistence of both VL and FM phases at the dashed line can not be discarded.}
\label{fig:DiagramaFase}
\end{figure}
We conclude that when $D\gg J$, the system undergoes a finite temperature phase transition with the same universal jump that the BKT transition, but now mediated by topologically trivial excitations.  In the extreme DM dominated regime, it is not difficult to show that the charge-charge correlation function decays exponentially with temperature \footnote{See supplementary information} and became less sensitive to charge positions, exactly in the same way that vortex-antivortex does, and therefore a decoupling of the neutral pair of topologically trivial excitations is expected at sufficiently high temperature, a phase diagram is shown in figure \ref{fig:DiagramaFase}.   The existence of a vortex-antivortex lattice and the iBKT transition it suffers, are the main results of this work. \\


{\it Acknowledgments.--}
We specially thank Daniel Cabra for very useful conversations.
We also thank Tom\'as Grigera, Rodolfo Borzi, Nicol\'as Grandi and Claudia Sc\'occola for useful discussions and suggestions.
This work was partially supported by  CONICET  (PIP  2015-813)  and  ANPCyT  (PICT2012-1724).

\pagebreak
.
\bibliography{references.bib}

\newcommand{\npb}{Nucl. Phys. B}\newcommand{\adv}{Adv.
  Phys.}\newcommand{\epl}{Europhys. Lett.}
\begin{thebibliography}{27}%
\makeatletter
\providecommand \@ifxundefined [1]{%
 \@ifx{#1\undefined}
}%
\providecommand \@ifnum [1]{%
 \ifnum #1\expandafter \@firstoftwo
 \else \expandafter \@secondoftwo
 \fi
}%
\providecommand \@ifx [1]{%
 \ifx #1\expandafter \@firstoftwo
 \else \expandafter \@secondoftwo
 \fi
}%
\providecommand \natexlab [1]{#1}%
\providecommand \enquote  [1]{``#1''}%
\providecommand \bibnamefont  [1]{#1}%
\providecommand \bibfnamefont [1]{#1}%
\providecommand \citenamefont [1]{#1}%
\providecommand \href@noop [0]{\@secondoftwo}%
\providecommand \href [0]{\begingroup \@sanitize@url \@href}%
\providecommand \@href[1]{\@@startlink{#1}\@@href}%
\providecommand \@@href[1]{\endgroup#1\@@endlink}%
\providecommand \@sanitize@url [0]{\catcode `\\12\catcode `\$12\catcode
  `\&12\catcode `\#12\catcode `\^12\catcode `\_12\catcode `\%12\relax}%
\providecommand \@@startlink[1]{}%
\providecommand \@@endlink[0]{}%
\providecommand \url  [0]{\begingroup\@sanitize@url \@url }%
\providecommand \@url [1]{\endgroup\@href {#1}{\urlprefix }}%
\providecommand \urlprefix  [0]{URL }%
\providecommand \Eprint [0]{\href }%
\providecommand \doibase [0]{http://dx.doi.org/}%
\providecommand \selectlanguage [0]{\@gobble}%
\providecommand \bibinfo  [0]{\@secondoftwo}%
\providecommand \bibfield  [0]{\@secondoftwo}%
\providecommand \translation [1]{[#1]}%
\providecommand \BibitemOpen [0]{}%
\providecommand \bibitemStop [0]{}%
\providecommand \bibitemNoStop [0]{.\EOS\space}%
\providecommand \EOS [0]{\spacefactor3000\relax}%
\providecommand \BibitemShut  [1]{\csname bibitem#1\endcsname}%
\let\auto@bib@innerbib\@empty
\bibitem [{\citenamefont {Berezinsky}(1970)}]{berezinsky1970}%
  \BibitemOpen
  \bibfield  {author} {\bibinfo {author} {\bibfnamefont {V.}~\bibnamefont
  {Berezinsky}},\ }\href@noop {} {\bibfield  {journal} {\bibinfo  {journal}
  {Zh. Eksp. Teor. Fiz.}\ }\textbf {\bibinfo {volume} {32}},\ \bibinfo {pages}
  {493} (\bibinfo {year} {1970})}\BibitemShut {NoStop}%
\bibitem [{\citenamefont {Berezinsky}(1972)}]{berezinsky1972}%
  \BibitemOpen
  \bibfield  {author} {\bibinfo {author} {\bibfnamefont {V.}~\bibnamefont
  {Berezinsky}},\ }\href@noop {} {\bibfield  {journal} {\bibinfo  {journal}
  {Zh. Eksp. Teor. Fiz.}\ }\textbf {\bibinfo {volume} {61}},\ \bibinfo {pages}
  {610} (\bibinfo {year} {1972})}\BibitemShut {NoStop}%
\bibitem [{\citenamefont {Kosterlitz}\ and\ \citenamefont
  {Thouless}(1973)}]{kosterlitz1973}%
  \BibitemOpen
  \bibfield  {author} {\bibinfo {author} {\bibfnamefont {J.~M.}\ \bibnamefont
  {Kosterlitz}}\ and\ \bibinfo {author} {\bibfnamefont {D.~J.}\ \bibnamefont
  {Thouless}},\ }\href@noop {} {\bibfield  {journal} {\bibinfo  {journal}
  {Journal of Physics C: Solid State Physics}\ }\textbf {\bibinfo {volume}
  {6}},\ \bibinfo {pages} {1181} (\bibinfo {year} {1973})}\BibitemShut
  {NoStop}%
\bibitem [{\citenamefont {Abrikosov}(1957)}]{abrikosov_57}%
  \BibitemOpen
  \bibfield  {author} {\bibinfo {author} {\bibfnamefont {A.~A.}\ \bibnamefont
  {Abrikosov}},\ }\href@noop {} {\bibfield  {journal} {\bibinfo  {journal}
  {Sov. Phys. JETP}\ }\textbf {\bibinfo {volume} {5}},\ \bibinfo {pages} {1174}
  (\bibinfo {year} {1957})}\BibitemShut {NoStop}%
\bibitem [{\citenamefont {Essmann}\ and\ \citenamefont
  {Tr{\"a}uble}(1967)}]{essmann1967}%
  \BibitemOpen
  \bibfield  {author} {\bibinfo {author} {\bibfnamefont {U.}~\bibnamefont
  {Essmann}}\ and\ \bibinfo {author} {\bibfnamefont {H.}~\bibnamefont
  {Tr{\"a}uble}},\ }\href@noop {} {\bibfield  {journal} {\bibinfo  {journal}
  {Physics Letters A}\ }\textbf {\bibinfo {volume} {24}},\ \bibinfo {pages}
  {526} (\bibinfo {year} {1967})}\BibitemShut {NoStop}%
\bibitem [{\citenamefont {Yarmchuk}\ \emph {et~al.}(1979)\citenamefont
  {Yarmchuk}, \citenamefont {Gordon},\ and\ \citenamefont
  {Packard}}]{yarmchuk1979}%
  \BibitemOpen
  \bibfield  {author} {\bibinfo {author} {\bibfnamefont {E.}~\bibnamefont
  {Yarmchuk}}, \bibinfo {author} {\bibfnamefont {M.}~\bibnamefont {Gordon}}, \
  and\ \bibinfo {author} {\bibfnamefont {R.}~\bibnamefont {Packard}},\
  }\href@noop {} {\bibfield  {journal} {\bibinfo  {journal} {Physical Review
  Letters}\ }\textbf {\bibinfo {volume} {43}},\ \bibinfo {pages} {214}
  (\bibinfo {year} {1979})}\BibitemShut {NoStop}%
\bibitem [{\citenamefont {Madison}\ \emph {et~al.}(2000)\citenamefont
  {Madison}, \citenamefont {Chevy}, \citenamefont {Wohlleben},\ and\
  \citenamefont {Dalibard}}]{madison2000}%
  \BibitemOpen
  \bibfield  {author} {\bibinfo {author} {\bibfnamefont {K.}~\bibnamefont
  {Madison}}, \bibinfo {author} {\bibfnamefont {F.}~\bibnamefont {Chevy}},
  \bibinfo {author} {\bibfnamefont {W.}~\bibnamefont {Wohlleben}}, \ and\
  \bibinfo {author} {\bibfnamefont {J.}~\bibnamefont {Dalibard}},\ }\href@noop
  {} {\bibfield  {journal} {\bibinfo  {journal} {Physical Review Letters}\
  }\textbf {\bibinfo {volume} {84}},\ \bibinfo {pages} {806} (\bibinfo {year}
  {2000})}\BibitemShut {NoStop}%
\bibitem [{\citenamefont {Abo-Shaeer}\ \emph {et~al.}(2001)\citenamefont
  {Abo-Shaeer}, \citenamefont {Raman}, \citenamefont {Vogels},\ and\
  \citenamefont {Ketterle}}]{abo2001}%
  \BibitemOpen
  \bibfield  {author} {\bibinfo {author} {\bibfnamefont {J.}~\bibnamefont
  {Abo-Shaeer}}, \bibinfo {author} {\bibfnamefont {C.}~\bibnamefont {Raman}},
  \bibinfo {author} {\bibfnamefont {J.}~\bibnamefont {Vogels}}, \ and\ \bibinfo
  {author} {\bibfnamefont {W.}~\bibnamefont {Ketterle}},\ }\href@noop {}
  {\bibfield  {journal} {\bibinfo  {journal} {Science}\ }\textbf {\bibinfo
  {volume} {292}},\ \bibinfo {pages} {476} (\bibinfo {year}
  {2001})}\BibitemShut {NoStop}%
\bibitem [{\citenamefont {Teitel}\ and\ \citenamefont
  {Jayaprakash}(1983{\natexlab{a}})}]{teitel_83a}%
  \BibitemOpen
  \bibfield  {author} {\bibinfo {author} {\bibfnamefont {S.}~\bibnamefont
  {Teitel}}\ and\ \bibinfo {author} {\bibfnamefont {C.}~\bibnamefont
  {Jayaprakash}},\ }\href@noop {} {\bibfield  {journal} {\bibinfo  {journal}
  {Physical Review B}\ }\textbf {\bibinfo {volume} {27}},\ \bibinfo {pages}
  {598} (\bibinfo {year} {1983}{\natexlab{a}})}\BibitemShut {NoStop}%
\bibitem [{\citenamefont {Teitel}\ and\ \citenamefont
  {Jayaprakash}(1983{\natexlab{b}})}]{teitel_83b}%
  \BibitemOpen
  \bibfield  {author} {\bibinfo {author} {\bibfnamefont {S.}~\bibnamefont
  {Teitel}}\ and\ \bibinfo {author} {\bibfnamefont {C.}~\bibnamefont
  {Jayaprakash}},\ }\href@noop {} {\bibfield  {journal} {\bibinfo  {journal}
  {Physical Review Letters}\ }\textbf {\bibinfo {volume} {51}},\ \bibinfo
  {pages} {1999} (\bibinfo {year} {1983}{\natexlab{b}})}\BibitemShut {NoStop}%
\bibitem [{\citenamefont {Teitel}(2013)}]{teitel2013}%
  \BibitemOpen
  \bibfield  {author} {\bibinfo {author} {\bibfnamefont {S.}~\bibnamefont
  {Teitel}},\ }in\ \href@noop {} {\emph {\bibinfo {booktitle} {40 Years of
  Berezinskii--Kosterlitz--Thouless Theory}}}\ (\bibinfo  {publisher} {World
  Scientific},\ \bibinfo {year} {2013})\ pp.\ \bibinfo {pages}
  {201--235}\BibitemShut {NoStop}%
\bibitem [{\citenamefont {Yang}\ \emph {et~al.}(2018)\citenamefont {Yang},
  \citenamefont {Boulle}, \citenamefont {Cros}, \citenamefont {Fert},\ and\
  \citenamefont {Chshiev}}]{yang18}%
  \BibitemOpen
  \bibfield  {author} {\bibinfo {author} {\bibfnamefont {H.}~\bibnamefont
  {Yang}}, \bibinfo {author} {\bibfnamefont {O.}~\bibnamefont {Boulle}},
  \bibinfo {author} {\bibfnamefont {V.}~\bibnamefont {Cros}}, \bibinfo {author}
  {\bibfnamefont {A.}~\bibnamefont {Fert}}, \ and\ \bibinfo {author}
  {\bibfnamefont {M.}~\bibnamefont {Chshiev}},\ }\href@noop {} {\bibfield
  {journal} {\bibinfo  {journal} {Scientific Reports}\ }\textbf {\bibinfo
  {volume} {8}},\ \bibinfo {pages} {1} (\bibinfo {year} {2018})}\BibitemShut
  {NoStop}%
\bibitem [{\citenamefont {Nembach}\ \emph {et~al.}(2015)\citenamefont
  {Nembach}, \citenamefont {Shaw}, \citenamefont {Weiler}, \citenamefont
  {Ju{\'e}},\ and\ \citenamefont {Silva}}]{nembach15}%
  \BibitemOpen
  \bibfield  {author} {\bibinfo {author} {\bibfnamefont {H.~T.}\ \bibnamefont
  {Nembach}}, \bibinfo {author} {\bibfnamefont {J.~M.}\ \bibnamefont {Shaw}},
  \bibinfo {author} {\bibfnamefont {M.}~\bibnamefont {Weiler}}, \bibinfo
  {author} {\bibfnamefont {E.}~\bibnamefont {Ju{\'e}}}, \ and\ \bibinfo
  {author} {\bibfnamefont {T.~J.}\ \bibnamefont {Silva}},\ }\href@noop {}
  {\bibfield  {journal} {\bibinfo  {journal} {Nature Physics}\ }\textbf
  {\bibinfo {volume} {11}},\ \bibinfo {pages} {825} (\bibinfo {year}
  {2015})}\BibitemShut {NoStop}%
\bibitem [{\citenamefont {Luo}\ \emph {et~al.}(2019)\citenamefont {Luo},
  \citenamefont {Zhang},\ and\ \citenamefont {Liu}}]{luo19}%
  \BibitemOpen
  \bibfield  {author} {\bibinfo {author} {\bibfnamefont {H.-B.}\ \bibnamefont
  {Luo}}, \bibinfo {author} {\bibfnamefont {H.-B.}\ \bibnamefont {Zhang}}, \
  and\ \bibinfo {author} {\bibfnamefont {J.~P.}\ \bibnamefont {Liu}},\
  }\href@noop {} {\bibfield  {journal} {\bibinfo  {journal} {npj Computational
  Materials}\ }\textbf {\bibinfo {volume} {5}},\ \bibinfo {pages} {1} (\bibinfo
  {year} {2019})}\BibitemShut {NoStop}%
\bibitem [{Note1()}]{Note1}%
  \BibitemOpen
  \bibinfo {note} {The FFXY models studied in these works, have an uniform flux
  on each plaquette with $f_c=\protect \frac {\phi _c}{\phi _0}$, where $\phi
  _0$ is the flux quanta and $\phi _c$ is the flux on each plaquette, constant
  all along the sample. In the case studied here, there is no such flux, of
  course, but the analogy is immediate, and it would corresponds to an
  alternating flux through neighbouring plaquettes. So that, the $\protect
  \mathbb {Z}_2$ symmetry that in the FFXY model is spostaneously broken, is
  explicitly broken here}\BibitemShut {NoStop}%
\bibitem [{Note2()}]{Note2}%
  \BibitemOpen
  \bibinfo {note} {By proper modulation of DM interaction, lattices of vortices
  of arbitrary sizes can be built with the same tessellation
  technique}\BibitemShut {NoStop}%
\bibitem [{Note3()}]{Note3}%
  \BibitemOpen
  \bibinfo {note} {In this discrete context we will call a vortex, centered in
  a given plaquette, to every configuration of the order parameter $\theta
  (\protect \mathbf {r_i})$ that, when circulating around the center of such
  plaquette, accumulates an integer times of $2\pi $'s, in steps smaller than
  $\pi $. A minimal vortex, then, is a vortex of the minimal size, i.e. a
  vortex of the size of the plaquette. This kind of vortices has been
  considered widely in the literature (see, for example, Ref.~\protect
  \rev@citealpnum {teitel2013} and references therein).}\BibitemShut {Stop}%
\bibitem [{Note4()}]{Note4}%
  \BibitemOpen
  \bibinfo {note} {Both configurations have the same $U(1)$ manifold
  degeneracy}\BibitemShut {NoStop}%
\bibitem [{\citenamefont {Savit}(1978)}]{savit78}%
  \BibitemOpen
  \bibfield  {author} {\bibinfo {author} {\bibfnamefont {R.}~\bibnamefont
  {Savit}},\ }\href@noop {} {\bibfield  {journal} {\bibinfo  {journal}
  {Physical Review B}\ }\textbf {\bibinfo {volume} {17}},\ \bibinfo {pages}
  {1340} (\bibinfo {year} {1978})}\BibitemShut {NoStop}%
\bibitem [{\citenamefont {Abramowitz}\ and\ \citenamefont
  {Stegun}(1948)}]{abramowitz48}%
  \BibitemOpen
  \bibfield  {author} {\bibinfo {author} {\bibfnamefont {M.}~\bibnamefont
  {Abramowitz}}\ and\ \bibinfo {author} {\bibfnamefont {I.~A.}\ \bibnamefont
  {Stegun}},\ }\href@noop {} {\emph {\bibinfo {title} {Handbook of mathematical
  functions with formulas, graphs, and mathematical tables}}},\ Vol.~\bibinfo
  {volume} {55}\ (\bibinfo  {publisher} {US Government printing office},\
  \bibinfo {year} {1948})\BibitemShut {NoStop}%
\bibitem [{\citenamefont {Schakel}(2008)}]{schakel}%
  \BibitemOpen
  \bibfield  {author} {\bibinfo {author} {\bibfnamefont {A.~M.}\ \bibnamefont
  {Schakel}},\ }\href@noop {} {\emph {\bibinfo {title} {Boulevard of broken
  symmetries: effective field theories of condensed matter}}}\ (\bibinfo
  {publisher} {World Scientific Publishing Company},\ \bibinfo {year}
  {2008})\BibitemShut {NoStop}%
\bibitem [{\citenamefont {Jos{\'e}}\ \emph {et~al.}(1977)\citenamefont
  {Jos{\'e}}, \citenamefont {Kadanoff}, \citenamefont {Kirkpatrick},\ and\
  \citenamefont {Nelson}}]{jose1977renormalization}%
  \BibitemOpen
  \bibfield  {author} {\bibinfo {author} {\bibfnamefont {J.~V.}\ \bibnamefont
  {Jos{\'e}}}, \bibinfo {author} {\bibfnamefont {L.~P.}\ \bibnamefont
  {Kadanoff}}, \bibinfo {author} {\bibfnamefont {S.}~\bibnamefont
  {Kirkpatrick}}, \ and\ \bibinfo {author} {\bibfnamefont {D.~R.}\ \bibnamefont
  {Nelson}},\ }\href@noop {} {\bibfield  {journal} {\bibinfo  {journal}
  {Physical Review B}\ }\textbf {\bibinfo {volume} {16}},\ \bibinfo {pages}
  {1217} (\bibinfo {year} {1977})}\BibitemShut {NoStop}%
\bibitem [{Note5()}]{Note5}%
  \BibitemOpen
  \bibinfo {note} {Since the system only presents minimal vortices, $|n|>1$ can
  not be distinguished from $|n|=1$ and excitations with $|n|>1$ have no
  physical meaning and are not considered. Nevertheless, numerically we observe
  that as temperature rises, vortex density only decrease}\BibitemShut
  {NoStop}%
\bibitem [{\citenamefont {Ohta}\ and\ \citenamefont
  {Jasnow}(1979)}]{ohta1979xy}%
  \BibitemOpen
  \bibfield  {author} {\bibinfo {author} {\bibfnamefont {T.}~\bibnamefont
  {Ohta}}\ and\ \bibinfo {author} {\bibfnamefont {D.}~\bibnamefont {Jasnow}},\
  }\href@noop {} {\bibfield  {journal} {\bibinfo  {journal} {Physical Review
  B}\ }\textbf {\bibinfo {volume} {20}},\ \bibinfo {pages} {139} (\bibinfo
  {year} {1979})}\BibitemShut {NoStop}%
\bibitem [{\citenamefont {Fisher}\ \emph {et~al.}(1973)\citenamefont {Fisher},
  \citenamefont {Barber},\ and\ \citenamefont {Jasnow}}]{fisher1973helicity}%
  \BibitemOpen
  \bibfield  {author} {\bibinfo {author} {\bibfnamefont {M.~E.}\ \bibnamefont
  {Fisher}}, \bibinfo {author} {\bibfnamefont {M.~N.}\ \bibnamefont {Barber}},
  \ and\ \bibinfo {author} {\bibfnamefont {D.}~\bibnamefont {Jasnow}},\
  }\href@noop {} {\bibfield  {journal} {\bibinfo  {journal} {Physical Review
  A}\ }\textbf {\bibinfo {volume} {8}},\ \bibinfo {pages} {1111} (\bibinfo
  {year} {1973})}\BibitemShut {NoStop}%
\bibitem [{\citenamefont {Weber}\ and\ \citenamefont
  {Minnhagen}(1988)}]{weber1988monte}%
  \BibitemOpen
  \bibfield  {author} {\bibinfo {author} {\bibfnamefont {H.}~\bibnamefont
  {Weber}}\ and\ \bibinfo {author} {\bibfnamefont {P.}~\bibnamefont
  {Minnhagen}},\ }\href@noop {} {\bibfield  {journal} {\bibinfo  {journal}
  {Physical Review B}\ }\textbf {\bibinfo {volume} {37}},\ \bibinfo {pages}
  {5986} (\bibinfo {year} {1988})}\BibitemShut {NoStop}%
\bibitem [{Note6()}]{Note6}%
  \BibitemOpen
  \bibinfo {note} {See supplementary information}\BibitemShut {NoStop}%
\end{thebibliography}%


\newcommand{\npb}{Nucl. Phys. B}\newcommand{\adv}{Adv.
  Phys.}\newcommand{\epl}{Europhys. Lett.}
\begin{thebibliography}{5}%
\makeatletter
\providecommand \@ifxundefined [1]{%
 \@ifx{#1\undefined}
}%
\providecommand \@ifnum [1]{%
 \ifnum #1\expandafter \@firstoftwo
 \else \expandafter \@secondoftwo
 \fi
}%
\providecommand \@ifx [1]{%
 \ifx #1\expandafter \@firstoftwo
 \else \expandafter \@secondoftwo
 \fi
}%
\providecommand \natexlab [1]{#1}%
\providecommand \enquote  [1]{``#1''}%
\providecommand \bibnamefont  [1]{#1}%
\providecommand \bibfnamefont [1]{#1}%
\providecommand \citenamefont [1]{#1}%
\providecommand \href@noop [0]{\@secondoftwo}%
\providecommand \href [0]{\begingroup \@sanitize@url \@href}%
\providecommand \@href[1]{\@@startlink{#1}\@@href}%
\providecommand \@@href[1]{\endgroup#1\@@endlink}%
\providecommand \@sanitize@url [0]{\catcode `\\12\catcode `\$12\catcode
  `\&12\catcode `\#12\catcode `\^12\catcode `\_12\catcode `\%12\relax}%
\providecommand \@@startlink[1]{}%
\providecommand \@@endlink[0]{}%
\providecommand \url  [0]{\begingroup\@sanitize@url \@url }%
\providecommand \@url [1]{\endgroup\@href {#1}{\urlprefix }}%
\providecommand \urlprefix  [0]{URL }%
\providecommand \Eprint [0]{\href }%
\providecommand \doibase [0]{http://dx.doi.org/}%
\providecommand \selectlanguage [0]{\@gobble}%
\providecommand \bibinfo  [0]{\@secondoftwo}%
\providecommand \bibfield  [0]{\@secondoftwo}%
\providecommand \translation [1]{[#1]}%
\providecommand \BibitemOpen [0]{}%
\providecommand \bibitemStop [0]{}%
\providecommand \bibitemNoStop [0]{.\EOS\space}%
\providecommand \EOS [0]{\spacefactor3000\relax}%
\providecommand \BibitemShut  [1]{\csname bibitem#1\endcsname}%
\let\auto@bib@innerbib\@empty
\bibitem [{\citenamefont {Abramowitz}\ and\ \citenamefont
  {Stegun}(1948)}]{abramowitz48}%
  \BibitemOpen
  \bibfield  {author} {\bibinfo {author} {\bibfnamefont {M.}~\bibnamefont
  {Abramowitz}}\ and\ \bibinfo {author} {\bibfnamefont {I.~A.}\ \bibnamefont
  {Stegun}},\ }\href@noop {} {\emph {\bibinfo {title} {Handbook of mathematical
  functions with formulas, graphs, and mathematical tables}}},\ Vol.~\bibinfo
  {volume} {55}\ (\bibinfo  {publisher} {US Government printing office},\
  \bibinfo {year} {1948})\BibitemShut {NoStop}%
\bibitem [{Note1()}]{Note1}%
  \BibitemOpen
  \bibinfo {note} {Supplemental material; cumulant functions}\BibitemShut
  {NoStop}%
\bibitem [{\citenamefont {Schakel}(2008)}]{schakel}%
  \BibitemOpen
  \bibfield  {author} {\bibinfo {author} {\bibfnamefont {A.~M.}\ \bibnamefont
  {Schakel}},\ }\href@noop {} {\emph {\bibinfo {title} {Boulevard of broken
  symmetries: effective field theories of condensed matter}}}\ (\bibinfo
  {publisher} {World Scientific Publishing Company},\ \bibinfo {year}
  {2008})\BibitemShut {NoStop}%
\bibitem [{\citenamefont {Ohta}\ and\ \citenamefont
  {Jasnow}(1979)}]{ohta1979xy}%
  \BibitemOpen
  \bibfield  {author} {\bibinfo {author} {\bibfnamefont {T.}~\bibnamefont
  {Ohta}}\ and\ \bibinfo {author} {\bibfnamefont {D.}~\bibnamefont {Jasnow}},\
  }\href@noop {} {\bibfield  {journal} {\bibinfo  {journal} {Physical Review
  B}\ }\textbf {\bibinfo {volume} {20}},\ \bibinfo {pages} {139} (\bibinfo
  {year} {1979})}\BibitemShut {NoStop}%
\bibitem [{\citenamefont {Jos{\'e}}\ \emph {et~al.}(1977)\citenamefont
  {Jos{\'e}}, \citenamefont {Kadanoff}, \citenamefont {Kirkpatrick},\ and\
  \citenamefont {Nelson}}]{jose1977renormalization}%
  \BibitemOpen
  \bibfield  {author} {\bibinfo {author} {\bibfnamefont {J.~V.}\ \bibnamefont
  {Jos{\'e}}}, \bibinfo {author} {\bibfnamefont {L.~P.}\ \bibnamefont
  {Kadanoff}}, \bibinfo {author} {\bibfnamefont {S.}~\bibnamefont
  {Kirkpatrick}}, \ and\ \bibinfo {author} {\bibfnamefont {D.~R.}\ \bibnamefont
  {Nelson}},\ }\href@noop {} {\bibfield  {journal} {\bibinfo  {journal}
  {Physical Review B}\ }\textbf {\bibinfo {volume} {16}},\ \bibinfo {pages}
  {1217} (\bibinfo {year} {1977})}\BibitemShut {NoStop}%
\end{thebibliography}%

%
%
\end{document}